# An Orbital Interpretation of Decreasing Galaxy Velocity–Distance Relations in Systems Dominated by Giant Elliptical Galaxies

.MELATI Rabia[1], LARBAOUI Takia[2], BENZIDANE Mohammed Ridha[3] and LOUNIS Mourad [4]

*[1-2] Physics Department, Abdelhamid Ibn Badis University,*
*Route Natoinale N 11, Kharouba, 27000 Mostaganem,Algeria.*
*rabia.melati@univ-mosta.dz*
*[3]Elaboration and Characterization Physico-Mechanical and Metallurgical of Materials Laboratory (ECP3M), Abdelhamid Ibn Badis University-Mostaganem.*
*Route Nationale N°11, Kharrouba, 27000, Mostaganem, Algeria.*
*[4] Physics Department, University of Relizane*
*Cité Bourmadia, W. Relizane PB 48000, W. Relizane*

**Abstract:**

The persistence of approximately flat rotation curves observed in many galaxies, is commonly interpreted as evidence for extended dark matter halos or as an indication of modified gravitational laws at low accelerations.

In this work, we investigate an alternative orbital framework for the large-scale kinematics of galaxy systems dominated by giant elliptical galaxies. We propose that selected non-active companion galaxies form gravitationally bound orbital systems around a dominant central elliptical galaxy, conceptually analogous to the Solar System

Using galaxy group catalogs together with data from the NASA/IPAC Extragalactic Database, we analyze three independent systems centered on M87, M60, and NGC 1399. Candidate companions are selected according to common observational and dynamical criteria, including their location within the adopted search volume, approximate spatial coplanarity, and consistency of the effective gravitational parameter ($G_{eff}$).

For the three systems, the resulting velocity–distance relations exhibit clear decreasing trends with increasing distance from the central elliptical galaxy, qualitatively consistent with Keplerian-like orbital dynamics. The derived mean effective gravitational parameters are ($1.78935 \text{ x}10^{-6}$ $m^3 \cdot kg^{-1} \cdot s^{-2}$, $1.4276 \times 10^{-6}$ $m^3$ $kg^{-1}$ $s^{-2}$ and $G_{eff}=1.38404\times 10^{-6}$ $m^3$ $kg^{-1}$ $s^{-2}$) for M87, M60, and NGC 1399, respectively. Despite the independence of these systems, the values lie within a relatively narrow range, suggesting a recurring effective dynamical scale within the class of giant-elliptical-dominated systems investigated here. These quantities are interpreted as effective parameters of the proposed framework rather than as replacements for the universal Newtonian gravitational constant.

Taken together, the spatial and kinematic results suggest that the selected systems can be described as hierarchical orbital configurations dominated by central giant elliptical galaxies. The proposed framework is not intended as a universal alternative to ΛCDM or MOND, but rather as a complementary dynamical interpretation for a specific class of elliptical-dominated galaxy systems in which decreasing velocity–distance relations are observed.

**Keywords**: Galaxy dynamics; Giant elliptical galaxies; Orbital motion; Decreasing velocity–distance relations; Newtonian dynamics; Galaxy groups

## 1. INTRODUCTION

Galaxy kinematics has long provided a crucial testing ground for gravitational theory. Early observations of spiral galaxies revealed rotation curves that remain approximately flat well beyond the extent of their luminous disks (Rubin et al. 1980; Bosma 1981), in apparent tension with the Keplerian decline expected from the distribution of visible matter alone. Within the standard cosmological framework, this discrepancy is commonly interpreted as evidence for massive, extended halos of non-baryonic dark matter surrounding galaxies (Blumenthal et al. 1984; Frenk & White 2012). This interpretation is now widely adopted and supported by numerous observational and theoretical studies (Hoeneisen 2019; Rodrigues et al. 2023; Bañares-Hernández et al. 2023; Roper et al. 2023; Magee et al. 2023). While the ΛCDM (Lambda Cold Dark Matter) paradigm has been remarkably successful in describing observations on cosmological scales, including the cosmic microwave background and the large-scale distribution of matter, some tensions and open questions remain at galactic scales, notably concerning the central density profiles of dark matter halos and the observed diversity of galaxy rotation curves (Bullock & Boylan-Kolchin 2017).

An alternative explanation is provided by Modified Newtonian Dynamics MOND (Modified Newtonian Dynamics) (Milgrom 1983), which postulates a modification of Newtonian dynamics below a characteristic acceleration scale, ($a_0 \approx 1.2\times10^{-10}$ ms$^{-2}$). MOND successfully reproduces many galactic rotation curves without invoking dark matter, particularly in low-surface-brightness galaxies (Famaey & McGaugh 2012). More broadly, a variety of alternative approaches have been proposed, including modified gravity frameworks and emergent phenomena at large scales (Tortora et al. 2018; Ludwig 2021; Harada 2022; Lasenby et al. 2023; Jiqing 2023; Toxvaerd 2024). However, extending MOND consistently to galaxy clusters and to a complete cosmological framework remains an active area of research, particularly regarding the formation and evolution of large-scale structure.

Despite their conceptual differences, both ΛCDM-based and MOND-based interpretations have been extensively applied to the rotation-supported dynamics of disk galaxies. However, this framework may not directly capture the dynamical organization of all galaxy systems, particularly those characterized by a strong gravitational hierarchy and the presence of a dominant central galaxy. In such systems, the dynamical configuration may differ substantially from that of rotation-supported disks, raising the question of whether complementary descriptions of the observed kinematics may be appropriate.

In this work, we explore an interpretation applicable to such environments. We propose that, in galaxy systems dominated by giant elliptical galaxies, the kinematics of surrounding galaxies can also be described in terms of gravitationally bound orbital motion around a dominant central giant elliptical galaxy, rather than being interpreted solely through the framework of collective rotation. This perspective is motivated by the observed tendency of giant elliptical galaxies to occupy the dynamical centers of galaxy groups and clusters and to exert a dominant gravitational influence on their surroundings (Binggeli et al. 1987). In this context, companion galaxies may be viewed as constituents of a hierarchical orbital system, analogous in its dynamical organization—though not in scale—to the Solar System.

Within this framework, non-active companion galaxies may follow Keplerian-like orbital trajectories around the central giant elliptical galaxy, naturally leading to decreasing velocity–distance relations consistent with the predictions of classical Newtonian gravity. This interpretation does not require the introduction of additional mass components or modifications of gravitational laws within the systems considered here.

The aim of this study is twofold. First, we test the viability of this orbital reinterpretation by applying it to three well-characterized galaxy systems centered on giant elliptical galaxies—M87 and M60 in the Virgo Cluster, and NGC 1399 in the Fornax Cluster. Second, we investigate whether these systems exhibit comparable, system-specific effective gravitational parameters ($G_{eff}$), potentially providing new insight into the large-scale dynamics of hierarchical galaxy systems. By analyzing the spatial configuration and kinematics of non-active companion galaxies, we assess whether their motions are consistent with a Newtonian orbital framework and whether the resulting velocity–distance relations exhibit the predicted decreasing behavior. More broadly, this approach emphasizes the importance of geometric structure, gravitational hierarchy, and orbital organization in galaxy systems, providing a complementary perspective within the broader study of galactic dynamics.

## 2. THEORETICAL FRAMEWORK

### 2.1 Classical Gravitation

Under the assumption of Newtonian gravity, the orbital velocity of a test particle at a distance *r* from a spherically symmetric mass distribution is given:

$$v = \sqrt{\frac{G\,M(r)}{r}} \tag{1}$$

where M(r) denotes the mass enclosed within radius *r*, and *G* is the gravitational constant.

Beyond the region containing most of the system's mass, this expression predicts a Keplerian decline of the velocity proportional to $r^{-1/2}$, as expected for gravitational systems dominated by a central mass, such as the Solar System.

### 2.2 General Relativity in the Weak-Field Regime

General relativity provides a more complete description of gravity by interpreting it as the curvature of spacetime and becomes essential in strong gravitational fields or at relativistic velocities. However, on galactic scales, where gravitational fields are weak and typical velocities are non-relativistic, Newtonian gravity generally provides an excellent approximation for describing orbital motion.

In this context, the present study investigates whether a Newtonian orbital framework can consistently describe the kinematics of companion galaxies in systems dominated by a central giant elliptical galaxy, and whether such an approach naturally produces decreasing velocity–distance relations without requiring additional mass components within the adopted orbital framework.

## 3. METHODOLOGY

The methodology adopted in this study is based on a conceptual comparison between two distinct dynamical configurations: centrally dominated orbital systems, exemplified by the Solar System, and galaxy systems in which the surrounding galaxies are conventionally analyzed through their velocity–distance relations.

The objective is to assess whether a centrally dominated orbital framework can also describe galaxy systems composed of a dominant central giant elliptical galaxy and surrounding non-active companion galaxies. In this context, we investigate whether the motions of the companion galaxies can be interpreted as gravitationally bound orbital motion, naturally producing decreasing velocity–distance relations consistent with classical Newtonian gravity.

The conceptual framework underlying this methodology is developed in the following subsection.

### 3.1 Conceptual Comparison: Planetary and Galactic Dynamical Regimes

Figure 1 illustrates the conceptual contrast between the Keplerian decline of orbital velocity expected in a centrally dominated system and the approximately flat rotation curves commonly observed in spiral galaxies.

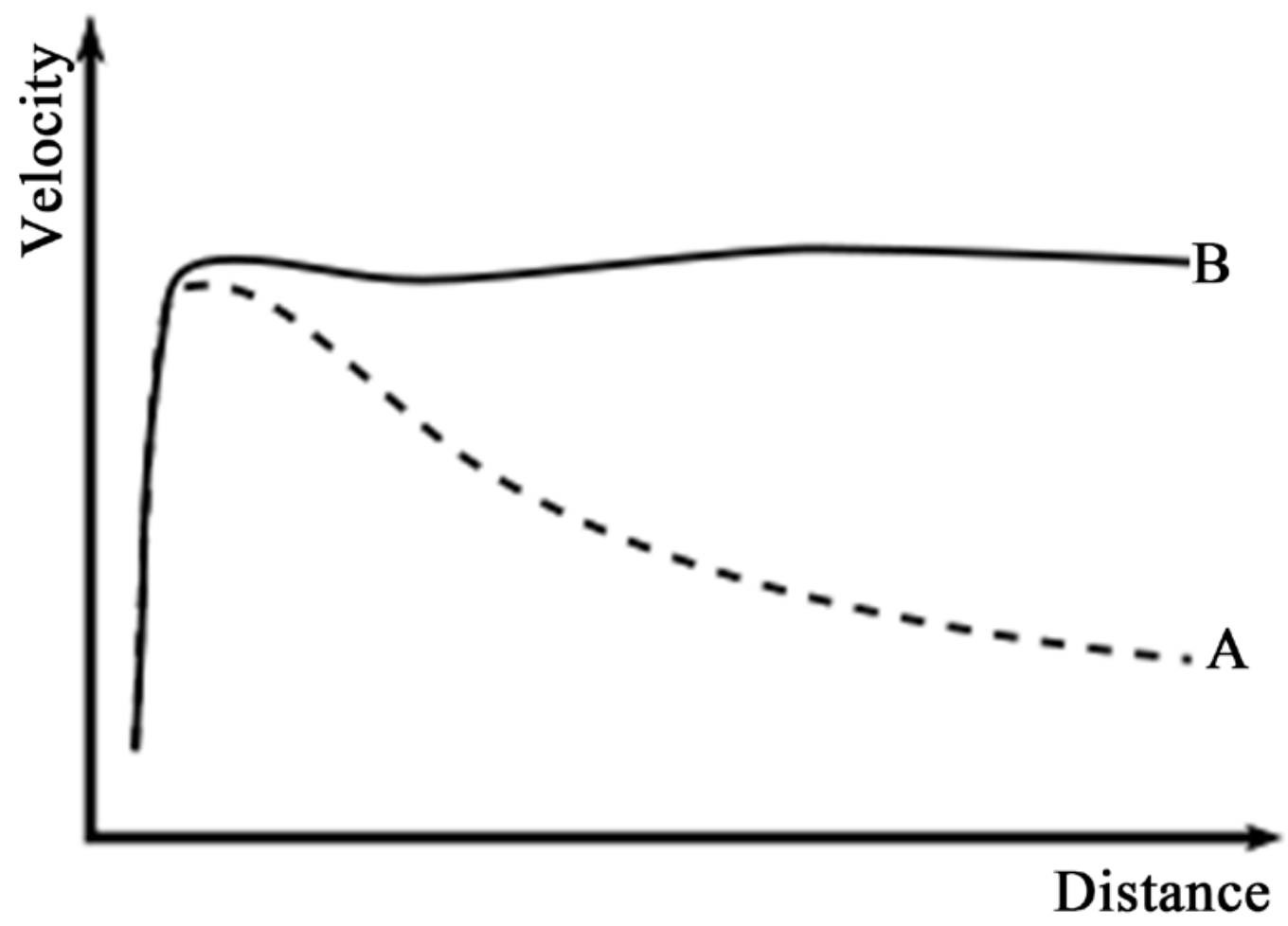

**Fig. 1.** Rotation curve of a typical spiral galaxy: (A) theoretical expectation and (B) observed profile

In planetary systems, each planet follows an individual gravitational orbit around a dominant central mass, with orbital velocities decreasing with distance according to Keplerian dynamics. Consequently, both orbital and angular velocities vary with radius, reflecting the centrally concentrated nature of the gravitational potential.

By contrast, in rotating galactic disks, stellar motions are not dominated by a single central mass but by the combined gravitational potential of the disk, bulge, and extended halo. In such systems, orbital velocities can remain approximately constant over a broad radial range, while angular velocities decrease with radius, indicating that the disk does not behave as a rigid body.

These differences reflect an important distinction between the two dynamical regimes: planetary systems are governed primarily by centrally dominated Keplerian orbits, whereas galactic disks are characterized by motion within an extended mass distribution. Consequently, the direct application of Keplerian expectations to disk-galaxy kinematics is not straightforward and may lead to apparently different velocity profiles.

This distinction suggests that environments dominated by a central giant elliptical galaxy may exhibit an orbital organization more closely analogous to that of planetary systems.

The following section develops this orbital framework and presents the physical arguments supporting its application to galaxy systems dominated by giant elliptical galaxies.

## 4. FOUNDATIONS OF THE NEW MODEL

The proposed framework is motivated by the physical characteristics and observed environments of giant elliptical galaxies located near the centers of galaxy groups and clusters. Their large masses, central positions, and extended gravitational influence make them natural candidates for representing the dominant gravitational component of selected galaxy systems. On this basis, the framework reinterprets selected non-active companion galaxies as members of an organized orbital configuration around a central giant elliptical galaxy.

### 4.1. Central gravitational dominance

Giant elliptical galaxies located at the centers of galaxy groups and clusters are among the most massive stellar systems in the local Universe (Dressler et al., 1980; Khosroshahi et al., 2006; Voevodkin, 2008). Their large stellar masses and central locations imply a substantial gravitational influence on their surrounding environments. Within the proposed framework, the central elliptical galaxy is therefore treated as the primary gravitational component governing the large-scale dynamics of the selected companion galaxies.

This configuration provides the physical basis for considering the surrounding non-active galaxies as gravitationally bound companions whose motions can, to first order, be described relative to the dominant central mass.

### 4.2. Mass hierarchy between the central galaxy and its companions

A second foundation of the proposed framework is the substantial mass contrast between the central giant elliptical galaxy and the selected companion galaxies. This hierarchy is conceptually analogous to the Sun–planet configuration, in which the central body dominates the gravitational potential while the orbiting bodies contribute comparatively little to the total mass.

Such a mass hierarchy provides a basis for adopting a central-mass approximation as a first-order description of the large-scale orbital dynamics of the selected companions.

### 4.3. Extended gravitational influence

The relevance of the central elliptical is not determined solely by its stellar mass but also by the spatial extent of its gravitational potential. The surrounding environment may therefore remain dynamically influenced by the central galaxy over scales substantially larger than its visible stellar body. Within the proposed framework, this extended gravitational influence provides a physical basis for investigating orbital motions of companion galaxies over megaparsec-scale separations.

### 4.4. Hierarchical orbital organization

Building on these considerations, the proposed framework describes a selected galaxy system as a hierarchical configuration in which a dominant central giant elliptical galaxy occupies the primary gravitational position and surrounding non-active galaxies constitute the associated orbital component.

The analogy with the Solar System is conceptual rather than structural. In planetary systems, the central mass dominates the gravitational potential and orbital velocities decrease with distance. The proposed framework investigates whether a comparable dynamical organization can be identified in selected giant-elliptical-dominated galaxy systems.

A further condition of this interpretation is that the selected companions form an approximately coplanar spatial configuration. This requirement is motivated by the nearly planar organization of planetary orbits and is used here to identify a geometrically coherent subset of companion galaxies. The quantitative assessment of this condition is presented separately in section 5.6.

Within this framework, the observed CMB-frame velocities are used as observational proxies for the large-scale orbital motions of the selected companion galaxies, while the effective gravitational parameter ($G_{eff}$) provides a system-specific measure of the dynamical scale inferred within the proposed framework.

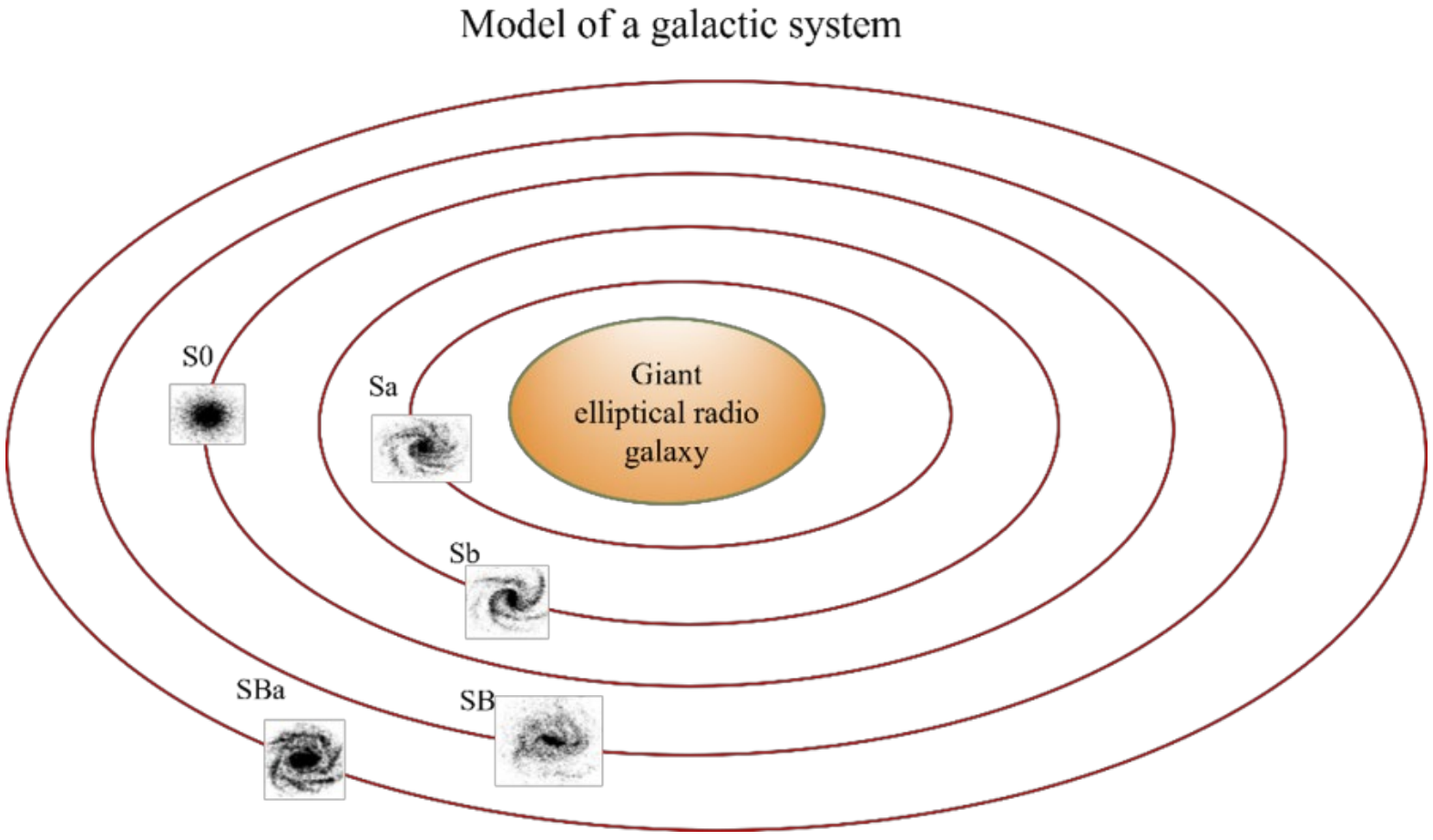

**Fig. 2.** Schematic representation of the proposed galaxy system model, highlighting its analogy with the Solar System.

## 5. METHODOLOGY FOR CONSTRUCTING VELOCITY–DISTANCE PROFILES IN THE ORBITAL FRAMEWORK

The fundamental assumption of the proposed framework is that the surrounding non-active companion galaxies can be regarded as individual companions gravitationally bound to a dominant central giant elliptical galaxy. Their motion is therefore interpreted as orbital motion around the central galaxy rather than as participation in a collective disk-like rotation.

To implement this framework, we developed a stepwise methodology designed to identify the companion galaxies of each galaxy system and to determine their dynamical properties in a consistent manner. The objective is to construct velocity–distance profiles and to assess whether the resulting orbital velocities follow the decreasing trend expected from Newtonian gravity.

To achieve this objective, the methodology consists of six sequential steps:

(i) Selection of candidate companion galaxies;

(ii) Application of the proposed membership criteria;

(iii) Compilation of CMB-corrected velocity measurements;

(iv) Calculation of orbital distances from the central giant elliptical galaxy;

(v) Estimation of the effective gravitational parameter ($G_{eff}$) within the proposed orbital framework;

(vi) Verification of the approximate coplanarity of the selected companion galaxies.

Together, these steps ensure that only galaxies satisfying the physical, geometrical, and dynamical requirements of the proposed orbital framework are included in the analysis.

To evaluate the applicability of the proposed framework, we apply this methodology to three representative galaxy systems centered on giant elliptical galaxies: NGC 4486 (M87), NGC 4649 (M60), and NGC 1399. Located within the Virgo and Fornax clusters, these systems were selected because of the dominant gravitational role of their central galaxies and the availability of suitable companion galaxies satisfying the adopted selection criteria. These systems therefore provide suitable test cases for assessing the applicability of the proposed orbital framework. The following subsections describe each step of the methodology in detail.

### 5.1 Selection of candidate companion galaxies

The candidate companion galaxies of the systems centered on NGC 4486 (M87), NGC 4649 (M60), and NGC 1399 were identified using two complementary approaches. First, established group catalogs (Ferguson 1989; Garcia 1993; Mahtessian 1998) were consulted to ensure consistency with previously recognized memberships. Second, radial velocity–constrained cone searches were performed in the NASA/IPAC Extragalactic Database (NED). For each central elliptical galaxy, its redshift-independent metric distance was adopted as the reference distance. The cone opening angle was then adjusted such that the projected search volume extended to a maximum radius of 10 Mpc at this distance. This choice provides a uniform selection criterion across the three systems, while the 10 Mpc limit should be regarded solely as an observational constraint imposed by the adopted search geometry, not as a physical boundary of the galaxy system. Figure 3 illustrates the geometry of the search cones for the three systems.

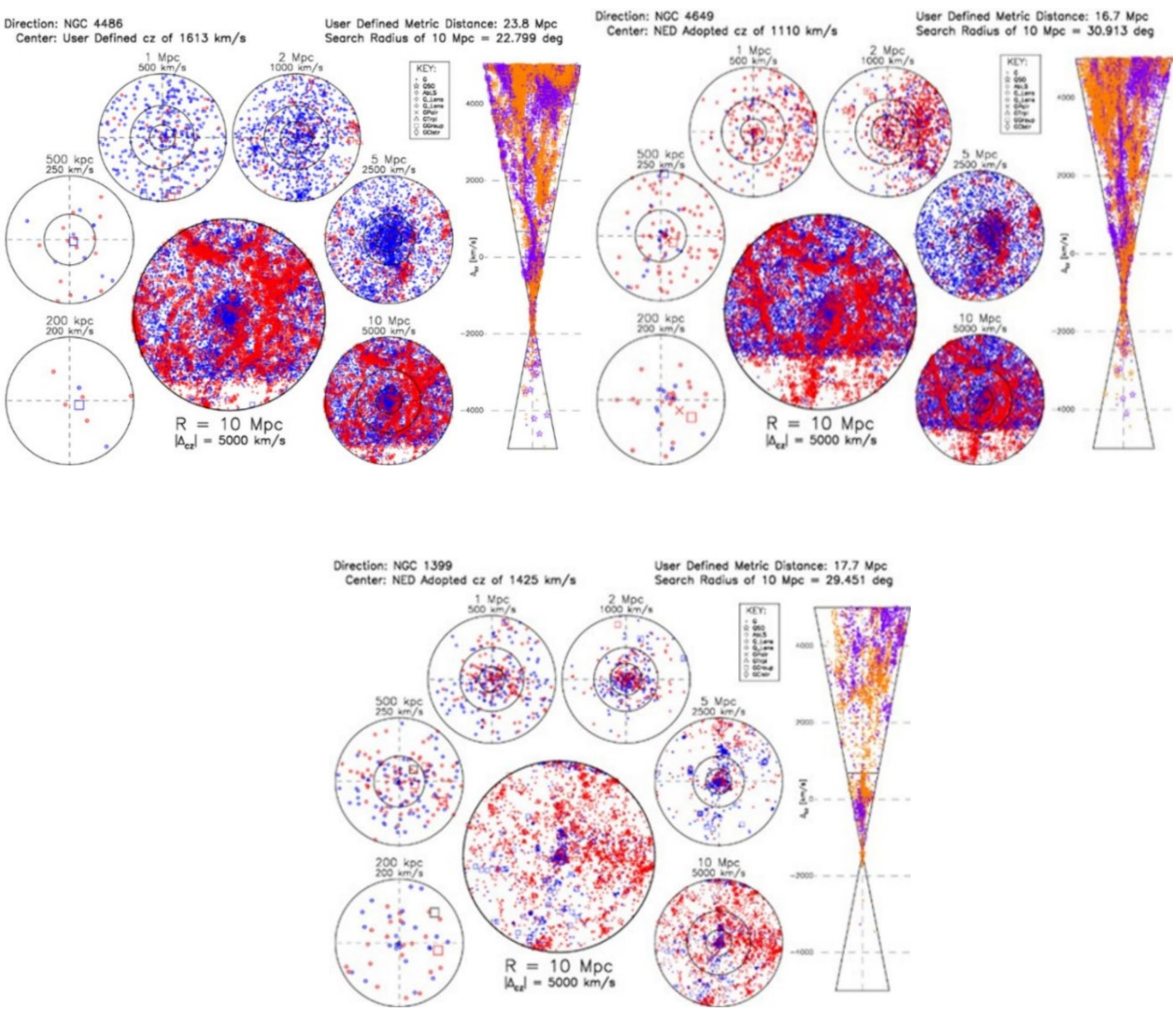

**Fig. 3.** View of the search cone around the galaxies NGC 4486, NGC 4649, and NGC 1399. based on radial data from NASA/IPAC.

### 5.2 Inclusion criteria

To ensure the robustness of the analysis, only galaxies satisfying a well-defined set of physical, geometrical, and dynamical criteria were retained as companion galaxies within the proposed orbital framework. Specifically:

(i) Galaxies must be non-active, in order to minimize potential internal kinematic distortions associated with nuclear or starburst activity.

(ii) They must lie within the adopted search volume.

(iii) They must satisfy the adopted approximate coplanarity criterion, quantified by the dimensionless parameter ($\eta = D_{\perp}/L$) (detailed in Section 5.6), with ($\eta \leq 0,10$). Galaxies satisfying ($\eta \leq 0,05$) are classified as strongly coplanar, while those satisfying ($0.05 < \eta \leq .10$) are classified as approximately coplanar.

(iv) Their effective gravitational parameters, ($G_{eff}$), estimated within the proposed orbital framework, must remain within a relatively narrow range, thereby providing a criterion for dynamical consistency among the selected companion galaxies.

### 5.3 Use of velocity data

The NASA/IPAC Extragalactic Database provides two velocity measurements for each galaxy: the heliocentric velocity (Helio) and the velocity relative to the cosmic microwave background (CMB). Since our model addresses large-scale galactic dynamics, we adopt the CMB-frame velocities, which are expressed in the CMB rest frame and provide a consistent tracer of systemic motion within the cosmological context. Although the CMB-frame recession velocity is not a direct measurement of the Newtonian orbital velocity, it is adopted here as a proxy for the large-scale motion of the companion galaxies relative to the central giant elliptical galaxy.

### 5.4 Orbital distance calculation

For each galaxy pair consisting of the central giant elliptical galaxy and a selected companion galaxy, the physical three-dimensional distance $D_{1-2}$ is computed using the law of cosines in three-dimensional space:

$$D_{1-2} = \sqrt{D_1^2 + D_2^2 - 2D_1 D_2 \cos\theta} \quad (2)$$

Where $D_1$ and $D_2$ are the CMB-corrected distances from Earth to the giant elliptical galaxy and the companion galaxy, respectively, and θ is the angle between their corresponding unit vectors in equatorial coordinates.

$$\theta = \cos^{-1}\left(\frac{\vec{r_1}.\vec{r_2}}{|\vec{r_1}|.|\vec{r_2}|}\right) \quad (3)$$

$\vec{r_1}$ and $\vec{r_2}$ are the unit vectors pointing from Earth toward the direction of the two galaxies.

$$\vec{r_1} = (\sin(\delta_1),\ \cos(\delta_1)\cos(\alpha_1),\ \cos(\delta_1)\sin(\alpha_1)) \quad (4)$$

$$\vec{r_2} = (\sin(\delta_2),\ \cos(\delta_2)\cos(\alpha_2),\ \cos(\delta_2)\sin(\alpha_2)) \quad (5)$$

δ and α are respectively the declination and right ascension, expressed in degrees

### 5.5 Estimation of effective gravitational parameter

Assuming orbital motion governed by Newtonian dynamics, we derive an effective gravitational parameter ($G_{eff}$) for each galaxy using:

$$G_{eff} = \frac{v^2.D_{1-2}}{M_{Ge}} \tag{6}$$

Where v is the galaxy's velocity relative to the CMB, interpreted in this framework as an orbital velocity around the central elliptical, and $M_{Ge}$ is the mass of the latter.

### 5.6 Coplanarity check

To assess whether the selected galaxies are approximately coplanar, their equatorial coordinates and CMB-frame Hubble distances were transformed into Cartesian coordinates (x,y,z). A reference plane was defined by three galaxies in each system, including the central elliptical galaxy. For each additional galaxy, the perpendicular distance to the reference plane was calculated as:

$$D_\perp = \frac{|ax_1 + by_1 + cz_1|}{\sqrt{a^2 + b^2 + c^2}} \tag{7}$$

To quantify the degree of spatial coplanarity, we define the dimensionless parameter $\eta = D_\perp/L$, where (L) is the characteristic three-dimensional extent of the system, defined as the maximum spatial separation between any two member galaxies.

$$L = \max(d_{i,j}) \quad \text{with } d_{i,j} = \sqrt{(x_i - x_j)^2 + (y_i - y_j)^2 + (z_i - z_j)^2} \tag{8}$$

We classify galaxies satisfying $\eta \leq 0.05$ as strongly coplanar, while those satisfying $0.05 < \eta \leq 0.10$ are classified as approximately coplanar. These thresholds are adopted as empirical criteria for the present analysis and are not intended to represent universal astrophysical limits. Galaxies with $\eta > 0.10$ are excluded from the coplanar subset used in the subsequent dynamical analysis.

## 6. APPLICATION OF THE PROPOSED FRAMEWORK TO SYSTEMS DOMINATED BY GIANT ELLIPTICAL GALAXIES

To assess the applicability and robustness of the proposed orbital framework, we analyze three representative galaxy systems dominated by giant elliptical galaxies: NGC 4486 (M87), NGC 4649 (M60), and NGC 1399, located in the Virgo and Fornax clusters. These galaxies are dynamically dominant within their local environments, making them suitable central objects for evaluating the proposed orbital framework. The selection of these galaxy systems was guided by three criteria:

(i) the presence of a dominant central giant elliptical galaxy with a strong gravitational influence;

(ii) the availability of a sufficient number of nearby non-active companion galaxies; and

(iii) reliable positional and kinematic data.

For each system, the methodology follows a structured sequence supported by the tabulated results (Tables 1–9). First, the equatorial coordinates and Hubble distances (CMB) of the companion galaxies are compiled (Tables 1, 4, and 7). Second, CMB-corrected velocities, orbital distances from the central galaxy, and effective gravitational parameters are determined (Tables 2, 5, and 8). Third, the approximate coplanarity of the selected companion galaxies is assessed according to the procedure described in Section 5.6 (Tables 3,6, and 9). This integrated methodology provides the basis for constructing velocity–distance profiles, evaluating their consistency with Newtonian orbital dynamics, and assessing the applicability of the proposed orbital framework to the selected galaxy systems.

## 6.1. Galaxy System Centered on M87

### 6.1.1. General Properties of M87

M87 (NGC 4486) is a supergiant elliptical galaxy located at a mean distance of approximately 16.757 Mpc (NASA/IPAC NED). and widely regarded as the central and most massive galaxy of the Virgo Cluster. It is also the brightest galaxy in the cluster (Binggeli et al. 1987; Bird et al. 2010). M87 hosts a supermassive black hole at its core, directly imaged by the Event Horizon Telescope (EHT Collaboration 2019). Owing to its extreme mass, central position, and dominant gravitational influence, M87 provides an ideal testbed for investigating orbital interpretations of galaxy dynamics in group environments.

### 6.1.2. Selection of companion galaxies

Companion galaxies associated with the M87 system were identified using the group catalogs of Garcia (1993) and Mahtessian (1998), complemented by radial velocity–constrained cone searches in the NASA/IPAC Extragalactic Database (NED). The final selection retained non-active companion galaxies satisfying the inclusion criteria defined in Section 5, including membership within the adopted search volume, approximate spatial coplanarity, and consistency of the effective gravitational parameter within the proposed orbital framework.

Table 1 summarizes the equatorial coordinates (RA, DEC) and CMB-frame Hubble distances of the selected galaxies.

**Table 1:** Equatorial coordinates (RA, DEC) and Hubble distances (CMB) of the companion galaxies in the M87 galaxy system

| Galaxy name | RA (Deg) | RA (Deg) | Hubble distance (CMB) [Mpc] |
|---|---|---|---|
| M87 | 187.705930 | 12.391123 | 23.79 ± 1.70 |
| MESSIER 087 DW12 | 187.831533 | 12.611554 | 30,88 ± 2.29 |
| NGC 4478 | 187.572578 | 12.328664 | 24.76 ± 1.77 |
| NGC 4477 | 187.509159 | 13.636604 | 24.55 ± 1.75 |
| VPC 0766 | 187.693292 | 12.099083 | 23.52 ± 1.68 |
| SDSSJ123048.60+120242.8 | 187.702520 | 12.045229 | 23.37 ± 1.67 |

| VCC 1313 | 187.702175 | 12.045016 | 23.38 ± 1.67 |
|---|---|---|---|
| VCC 1307 | 187.685917 | 13.799444 | 23.11 ± 1.89 |
| NGC 4298 | 185.386527 | 14.606049 | 21.43 ± 1.54 |
| NGC 4302 | 185.426257 | 14.597524 | 21.04 ± 1.52 |
| NGC 4497 | 187.88553 | 11.624718 | 20.24 ± 1.46 |
| SDSSJ123038.62+153629.1 | 187.660950 | 15.608089 | 15.95 ± 1.29 |

Using the data presented in Table 1 together with Eqs. (2)–(6), we calculated the three-dimensional orbital distances ($D_{1-2}$) between M87 and each companion galaxy, as well as the corresponding effective gravitational parameters ($G_{eff}$). The results are presented in Table 2. The ($G_{eff}$) values span a relatively narrow range from 1.53746 x $10^{-6}$ $m^3 \cdot kg^{-1} \cdot s^{-2}$ to 1.91883 x $10^{-6}$ $m^3 \cdot kg^{-1} \cdot s^{-2}$, with a mean value of 1.78935 x$10^{-6}$ $m^3 \cdot kg^{-1} \cdot s^{-2}$. This relatively limited dispersion suggests a consistent dynamical behavior among the selected companion galaxies within the proposed orbital framework.

**Table 2:** CMB-frame velocities, three-dimensional orbital distances ($D_{1-2}$) and corresponding effective gravitational parameters ($G_{eff}$) for the M87 galaxy system.

| Galaxy name | CMB velocity (Km/s) | Distance $D_{1-2}$ (Mpc) | Effective Gravitational Parameter $G_{eff}$ ($m^3 kg^{-1} s^{-2}$) |
|---|---|---|---|
| MESSIER 087 DW12 | 2094 ± 39 | 11.2530947 | 1.88223E-06 |
| NGC 4478 | 1678 ± 23 | 14.31433814 | 1.53746E-06 |
| NGC 4477 | 1665 ± 23 | 17.4591868 | 1.84629E-06 |
| VPC 0766 | 1595 ± 23 | 18.6913504 | 1.81388E-06 |
| SDSSJ123048.60+120242.8 | 1584 ± 23 | 19.79196546 | 1.89429E-06 |
| VCC 1313 | 1585 ± 23 | 19.8005752 | 1.89751E-06 |
| VCC 1307 | 1567 ± 50 | 20.1617199 | 1.88848E-06 |
| NGC 4298 | 1453 ± 23 | 20.55500612 | 1.65537E-06 |
| NGC 4302 | 1427 ± 24 | 20.80613327 | 1.61617E-06 |
| NGC 4497 | 1372 ± 23 | 26.7226738 | 1.91883E-06 |
| SDSSJ123038.62+153629.1 | 1082 ± 35 | 38,79080791 | 1.73233E-06 |

To assess the spatial coplanarity of the selected companion galaxies, a reference plane (P1) was defined by the central galaxy M87 and two companion galaxies, NGC 4298 and SDSS J123038.62+153629.1, selected from Table 1 on purely geometrical grounds. These two galaxies were chosen solely to establish the reference plane and were not selected on the basis of their dynamical properties. Using their Cartesian coordinates, the resulting reference plane is given by ($P_1$: -1.27036x+4.26983y-6.57797z+17.63221=0).

The perpendicular distances ($D_\perp$) of the companion galaxies to this plane were then calculated following the procedure described in Section 5.6. For the M87 system, the characteristic three-dimensional extent is (L=14.9752) Mpc, defined as the maximum spatial separation between any two member galaxies. The normalized parameter ($\eta = D_\perp/L$) was subsequently used to quantify the degree of coplanarity. The results are presented in Table 3, together with the corresponding coplanarity classifications.

**Table3:** Perpendicular distances $D_\perp$ of the companion galaxies to the reference plane ($P_1$).

| **Galaxy** | **$D_\perp$ (Mpc)** | **$D_\perp/L$** | **Coplanaire à 10 %?** |
|---|---|---|---|
| M87 | 000 | 0.00 % | yes |
| MESSIER 087 DW12 | 0.7963 | 5.32 % | Yes |
| NGC 4478 | 0.0369 | 0.25 % | Yes |
| NGC 4477 | 0.4664 | 3.11 % | Yes |
| VPC 0766 | 0.1272 | 0.85 % | Yes |
| SDSSJ123048.60+120242.8 | 0.1567 | 1.05 % | Yes |
| VCC 1313 | 0.1560 | 1.04 % | Yes |
| VCC 1307 | 0.4014 | 2.68 % | Yes |
| NGC 4298 | 0.000 | 0.00 % | Yes |
| NGC 4302 | 0.0349 | 0.23 % | Yes |
| NGC 4497 | 0.5209 | 3.48 % | Yes |
| SDSSJ123038.62+153629.1 | 0.000 | 0.00 % | Yes |

All 12 galaxies of table 3 satisfy the adopted criterion ($\eta = D_\perp/L \leq 0.10$), indicating that the full sample is approximately coplanar with respect to the reference plane. Moreover, 11 of the 12 galaxies have ($\eta \leq 0.05$), corresponding to a strongly coplanar configuration, while only MESSIER 087 DW12 lies marginally above the 5% threshold, with ($\eta$ =5.32%).

### 6.1.3. Velocity–Distance relation and orbital interpretation

For the final M87 system sample, the velocity–distance relation was constructed (Figure 4) by plotting the CMB-frame velocities of the selected companion galaxies as a function of their distances ($D_{1\text{-}2}$) from the central elliptical galaxy. Within the proposed framework, these velocities are interpreted as tracers of the large-scale orbital motion of the companion galaxies around M87.

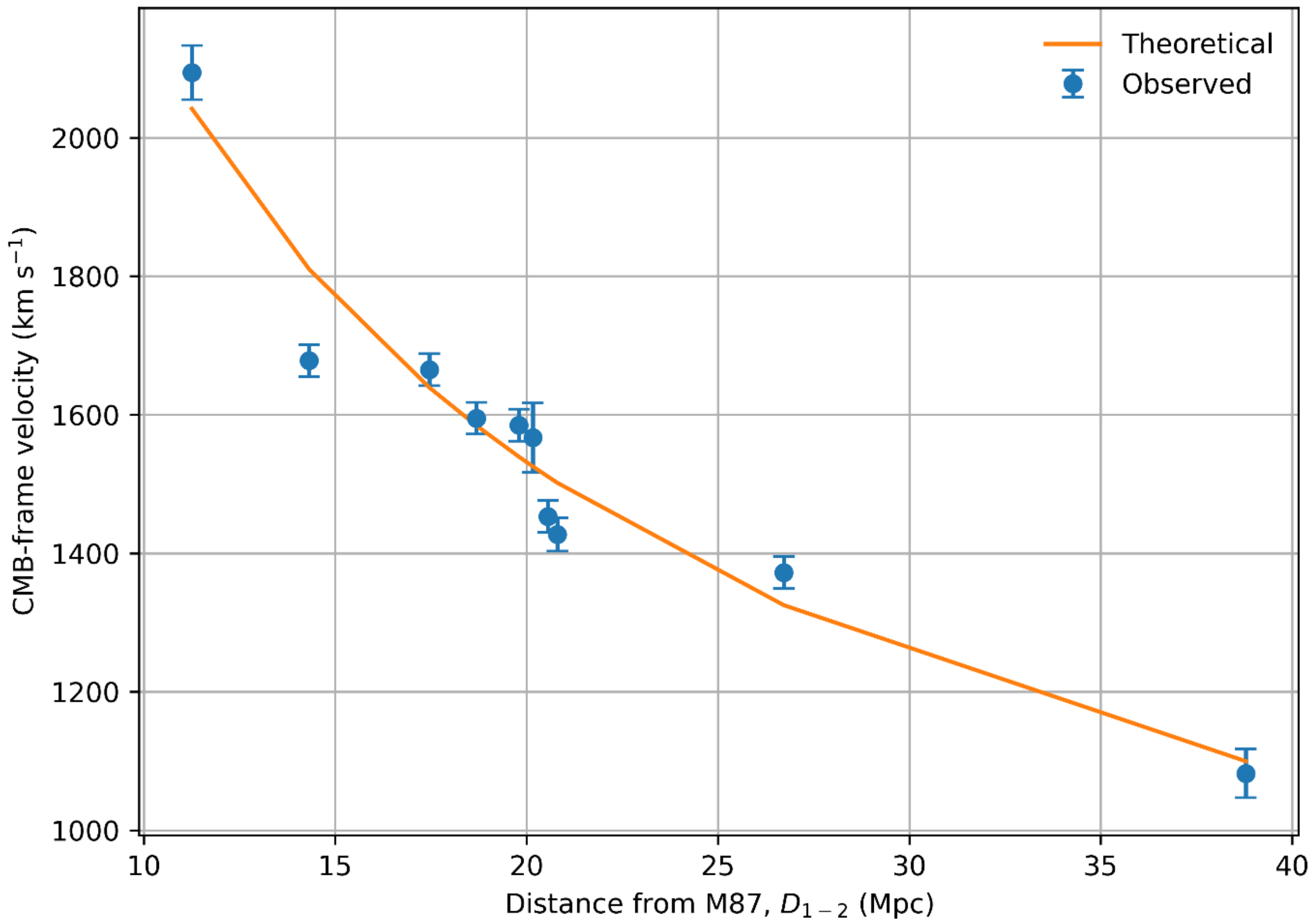

**Fig. 4.** Velocity–distance relation of the M87 system

### 6.1.3.1 Interpretation of the velocity–distance relation of the M87 system

Figure 4 shows the variation of the CMB-frame velocity as a function of the distance from M87 ($D_{1\text{-}2}$) for the galaxies identified as members of the proposed M87 system. The observational data display a clear overall decrease in velocity with increasing distance from the central giant elliptical galaxy. Although a limited scatter is present among individual galaxies, the measured velocities define a well-organized declining trend over the entire distance range investigated. The associated uncertainties remain small compared with the total velocity variation, indicating that the observed behavior is robust and not dominated by measurement errors.

The theoretical Keplerian profile was calculated using the mean effective gravitational parameter ($G_{eff}$ = 1.78935 x $10^{-6}$ $m^3$ $kg^{-1}s^{-2}$), the photometric mass of M87 ($M_{phot}$(M87) = 8.09 x $10^{41}$ kg), and the same orbital distances ($D_{1\text{-}2}$) that were used to derive the individual values of ($G_{eff}$). This ensures that the observational and theoretical velocity profiles are established within a common dynamical framework. The theoretical curve reproduces the expected Keplerian decrease of orbital velocity with distance and follows the same global evolution as the observational data. The remaining deviations between individual galaxies and the theoretical profile are moderate and can reasonably be attributed to local dynamical effects, measurement uncertainties, line-of-sight projection, and the intrinsic three-dimensional distribution of the galaxies, rather than to a systematic departure from the proposed model.

The declining velocity profile obtained for the M87 system differs fundamentally from the nearly flat rotation curves commonly observed in spiral galaxies. Instead, it is consistent with a gravitationally dominated system in which the orbital motion of the surrounding galaxies is primarily governed by the baryonic mass of the central giant elliptical galaxy. This behavior is fully consistent with the conceptual framework proposed in this work, where giant elliptical galaxies act as the dominant gravitational centers of organized galactic systems and their surrounding non-active galaxies behave as orbiting companions. When considered together with the coplanar spatial distribution established for the same galaxies, the velocity–distance relation provides independent dynamical evidence supporting the existence of a coherent orbital structure around M87. Consequently, the M87 system constitutes a representative example illustrating the applicability of the proposed orbital model to giant elliptical galaxy systems.

### 6.2. Galaxy System Centered on M60

#### 6.2.1. General Properties of M60

M60 is a giant elliptical galaxy (Bridges 2006; Sharples 2006) located at a mean distance of approximately 16.702 Mpc (NASA/IPAC NED). It is among the most massive galaxies in the Virgo Cluster (Ferrarese et al. 2006) and is the third-brightest elliptical galaxy in the cluster (Mei et al. 2007; Mamon 2008). M60 hosts a supermassive black hole at its center (Shen 2010) and exhibits detectable X-ray emission, consistent with the presence of a weakly active galactic nucleus. Owing to its dominant gravitational influence within its local environment and the presence of several nearby non-active companion galaxies, M60 provides a suitable test case for investigating whether the kinematics of companion galaxies are consistent with the proposed orbital framework rather than with a collective disk-like rotation.

#### 6.2.2. Selection of Companion Galaxies and Dynamical Analysis

Applying the methodology described in Section 5, seven non-active companion galaxies were selected for the M60 system. Their equatorial coordinates (right ascension and declination) and CMB-frame Hubble distances are summarized in Table 4.

Using these data together with Eqs. (2)–(6), we calculated the three-dimensional orbital distances ($D_{1\text{-}2}$) between M60 and each selected companion galaxy, together with the corresponding effective gravitational parameters ($G_{eff}$). The results, presented in Table 5, range from ($1.27916 \times 10^{-6}$ to $1.5919 \times 10^{-6}\ m^3Kg^{-1}s^{-2}$), with a mean value of ($G_{eff} = 1.4276 \times 10^{-6}\ m^3Kg^{-1}s^{2}$). The relatively narrow dispersion of these values suggests a consistent dynamical behavior among the selected companion galaxies within the proposed orbital framework.

To assess the spatial coplanarity of the selected companion galaxies, a reference plane ($P_2$) was defined by the central galaxy M60 and two companion galaxies, NGC 4651 and NGC 4630, selected from Table 4 on purely geometrical grounds. These two galaxies were chosen solely to establish the reference plane and were not selected on the basis of their dynamical properties. The resulting reference plane is given by ($P_2$:-5.1500y-0.1225z+0.6554=0).

The perpendicular distances ($D_\perp$) of the selected galaxies to this plane, including the three galaxies used to define the reference plane, are presented in Table 6, the three defining galaxies have ($D_\perp=0$) by construction. For the final M60 system sample, the characteristic three-dimensional extent is (L=5.5405) Mpc, defined as the maximum spatial separation between any two member galaxies. The largest perpendicular distance, ($D_\perp$=0.1665 Mpc), corresponds to ($D_\perp/L$=0.0301), which is well below the adopted 10% threshold. Thus, all selected companion galaxies satisfy the adopted approximate coplanarity criterion.

Taken together, these results are consistent with those obtained for the M87 system, indicating that the same spatial and dynamical characteristics may recur in independent galaxy systems dominated by giant elliptical galaxies.

**Table 4 :** Equatorial coordinates (RA, Dec) and Hubble distances (CMB) of the companion galaxies in the M60 galaxy system

| Galaxy | RA (Deg) | RA (Deg) | Hubble distance (CMB) [Mpc] |
|---|---|---|---|
| M60 | 190.916564 | 11.552706 | 21.14 +/- 1.52 |
| NGC 4635 | 190.663503 | 19.945312 | 18.59 +/- 1.34 |
| NGC 4600 | 190.095664 | 3.117750 | 17.32 +/- 1.26 |
| VCC 1950 | 190.751479 | 16.087184 | 16.02 +/- 1.20 |
| NGC 4651 | 190.927625 | 16.393389 | 16.41 +/- 1.19 |
| NGC 4630 | 190.629706 | 3.960245 | 15.89 +/- 1.16 |
| IC 3647 | 190.221333 | 10.475364 | 14.18 ± 1.78 |
| VCC 1839 | 190.108008 | 4.050149 | 15.70 +/- 1.28 |

**Table 5.** CMB-frame velocities, three-dimensional orbital distances ($D_{1-2}$), and corresponding effective gravitational parameters ($G_{\text{eff}}$) for the M60 galaxy system.

| Galaxy | CMB velocity (Km/s) | Distance $D_{1-2}$ (Mpc) | Effective gravitational parameter $G_{eff}$ ($m^3 kg^{-1} s^{-2}$) |
|---|---|---|---|
| **NGC 4635** | 1260 +/- 21 | 19.5769813 | 1.38993E-06 |
| **NGC 4600** | 1174 +/- 24 | 21.63879 | 1.33375E-06 |
| **VCC 1950** | 1086 +/- 26 | 24.2527156 | 1.27916E-06 |
| **NGC 4651** | 1113 +/- 22 | 28.3720528 | 1.57176E-06 |
| **NGC 4630** | 1077 +/- 24 | 29.7154869 | 1.54141E-06 |
| **IC 3647** | 961 ± 73 | 31.1211526 | 1.28531E-06 |
| **VCC 1839** | 1064 +/- 36 | 31.4432391 | 1.5919E-06 |

**Table6 :** Perpendicular distances of the companion galaxies to the reference plane ($P_2$).

| **Galaxy** | **D⊥ (Mpc)** | **D⊥/L** | **Coplanaire à 10 %?** |
|---|---|---|---|
| M60 | 000 | 0.00 % | yes |
| NGC 4635 | 0.1223 | 2.21 % | Yes |
| NGC 4600 | 0.1665 | 3.01 % | Yes |
| VCC 1950 | 0.0424 | 0.77 % | Yes |
| NGC 4651 | 0.0000 | 0.00 % | Yes |
| NGC 4630 | 0.0000 | 0.00 % | Yes |
| IC 3647 | 0.1220 | 2.20 % | Yes |
| VCC 1839 | 0.1416 | 2.56 % | Yes |

### 6.2.3. Velocity–Distance Relation of the M60 system and Orbital Interpretation

For the final M60 galaxy sample, the velocity–distance relation was constructed (Figure 5) by plotting the CMB-frame velocities of the companion galaxies as a function of their distances ($D_{1\text{-}2}$) from the central elliptical galaxy. Within the proposed framework, these velocities are interpreted as tracers of the large-scale orbital motion of the companion galaxies around M60.

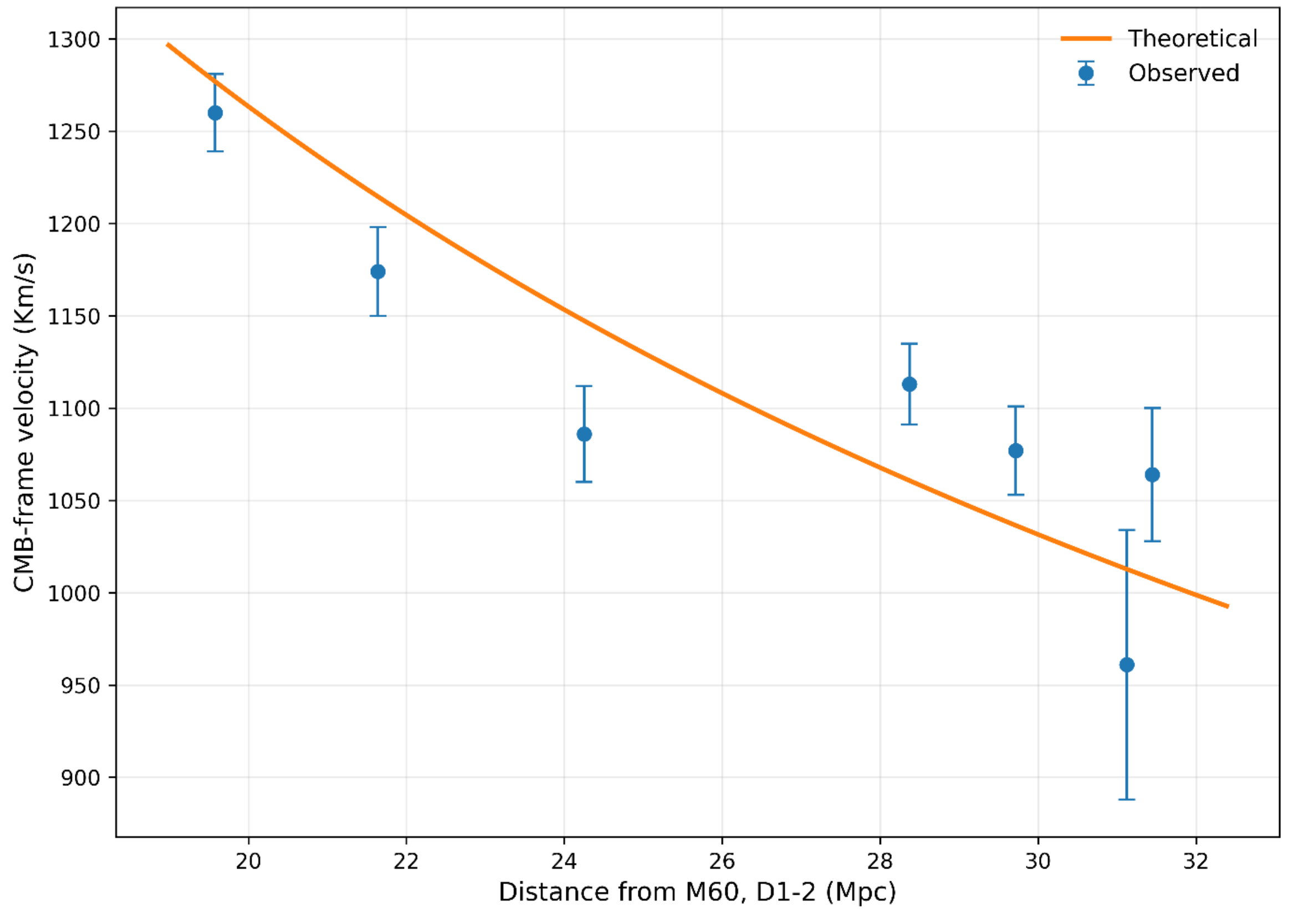

**Fig. 5.** Velocity–distance relation of the M60 system

#### 6.2.3.1. Interpretation of the velocity–distance relation of the M60 system

The observed velocities in Figure 5 show a clear overall decrease with increasing distance ($D_{1\text{-}2}$) despite a moderate scatter among individual measurements. The theoretical Keplerian-like profile, calculated using the mean effective gravitational parameter ($G_{eff}$ =1.4276 x$10^{-6}$ $m^3kg^{-1}s^{-2}$) and the photometric mass of M60, ($M_{phot}$ (M60) = 6.90×$10^{41}$ kg), exhibits the same overall declining behavior. Because the same orbital distances ($D_{1\text{-}2}$) were used to derive the individual ($G_{eff}$) values and to construct the theoretical profile, the comparison is internally consistent. The observed and theoretical profiles therefore show the same overall radial trend in velocity within the framework of the proposed model.

Together with the previously established coplanar configuration of the selected galaxies, the decreasing velocity–distance relation is consistent with the interpretation of M60 as the dominant central gravitational component of an organized galactic system. The result is qualitatively consistent with Keplerian-like orbital dynamics, although the limited number of galaxies and the scatter around the theoretical profile warrant a cautious interpretation.

## 6.3. Galaxy System Centered on NGC 1399

### 6.3.1. General Properties of NGC 1399

NGC 1399 is a giant elliptical galaxy of morphological type E0 (Killeen 1988), located at a mean distance of approximately 17.711 Mpc according to the NASA/IPAC Extragalactic Database. It is the central and dynamically dominant galaxy of the Fornax Cluster (Shurkin 2008), where it constitutes one of the most massive and luminous members. NGC 1399 hosts a supermassive black hole associated with an active galactic nucleus (Loewenstein et al. 2001; Graham 2008) and exhibits extended radio emission in the form of antiparallel jets and diffuse lobes (Killeen 1988). Owing to its dominant gravitational influence within the Fornax Cluster and the presence of several nearby non-active companion galaxies, NGC 1399 provides an appropriate test case for evaluating the proposed orbital framework in an environment distinct from the Virgo Cluster.

### 6.3.2. Selection of Companion Galaxies and Dynamical Analysis

Applying the methodology described in Section 5, a set of non-active companion galaxies was selected for the NGC 1399 galaxy system. Their equatorial coordinates (right ascension and declination) and Hubble distances (CMB), are summarized in Table 7.

Using these data together with Eqs. (2)–(6), we computed the three-dimensional orbital distances ($D_{1\text{-}2}$) between NGC 1399 and each companion galaxy, as well as the corresponding effective gravitational parameters ($G_{eff}$). The results, presented in Table 8, show values ranging from (1.20433×$10^{-6}$ to 1.58963×$10^{-6}$ $m^3$ $kg^{-1}$ $s^{-2}$), with a mean value of ($G_{eff}$=1.38404×$10^{-6}$ $m^3$ $kg^{-1}$ $s^{-2}$). As in the previous systems, the relatively narrow dispersion of $G_{eff}$ values indicates a high degree of dynamical consistency within the proposed orbital framework.

To investigate the spatial coplanarity of the selected companion galaxies, a reference plane, denoted as $P_3$, was constructed from the central galaxy NGC 1399 together with two companion galaxies, NGC 1387 and IC 1970, selected from the objects listed in Table7. This selection is based solely on geometrical considerations, since these two galaxies are used only as reference points for establishing the plane. From their cartesian coordinates, the corresponding plane is expressed as (x -0.61255 y +0.01347 z -1.11987= 0).

The perpendicular distance ($D_\perp$) from each of the other companion galaxies to this reference plane was subsequently determined, with the resulting values reported in Table 9. The degree of coplanarity was then quantified through the dimensionless parameter ($\eta = D_\perp/L$), following the criterion introduced in Section 5.6. Here, (L=15.6106) Mpc represents the maximum three-dimensional separation between any two member galaxies. Consequently, adopting the criterion ($\eta = D_\perp/L \leq 0.10$), the corresponding upper limit for the perpendicular distance is (0.10L=1.56106) Mpc.

The results of table 9 show that all eight galaxies satisfy the adopted coplanarity criterion, with ($D_\perp/L \leq 0.10$ (10%)). Moreover, all eight galaxies have ($D_\perp/L \leq 0.05$), indicating a strongly coplanar configuration. The largest relative deviation is found for NGC 1351 (4.31 %), while the three galaxies defining the reference plane, NGC 1399, NGC 1387, and IC 1970, have ($D_\perp=0$ Mpc) by construction.

**Table 7:** Equatorial coordinates (RA, DEC) and Hubble distances (CMB) of the companion galaxies in the NGC 1399 galaxy system

| Galaxy | RA (Deg) | DEC (Deg) | Hubble distance (CMB) [Mpc] |
|---|---|---|---|
| NGC 1399 | 54.620941 | -35.450657 | 19.65 ± 1.38 mv |
| NGC 1351 | 52.645790 | -34.854056 | 21.07 ± 1.48 |
| ESO 548- G 063 | 54.894917 | -20.014976 | 27.54 ± 1.94 |
| IC 1970 | 54.131333 | -43.956833 | 17.21 ± 1.21 |
| NGC 1377 | 54.162833 | -20.90225 | 24.61 ± 1.84 |
| NGC 1387 | 54.237641 | -35.506 | 17.82 ± 1.25 |
| NGC 1351A | 52.202877 | -35.177968 | 18.41 ± 1.29 |
| NGC 1406 | 54.847083 | -31.321417 | 14.37 ± 1.01 |

.**Table 8:** CMB-frame velocities, three-dimensional orbital distances ($D_{1-2}$), and corresponding effective gravitational parameters ($G_{\mathrm{eff}}$) for the NGC 1399 galaxy system.

| Galaxy name | CMB velocity (Km/s) | Distance $D_{1-2}$ (Mpc) | Gravitational parameter $G_{eff}$ ($m^3kg^{-1}$ $s^{-2}$) |
|---|---|---|---|
| **ESO 548- G 063** | 1868 ± 13 | 11.190094 | 1.55668E-06 |
| **NGC 1377** | 1669 ± 31 | 14.314314 | 1.58963E-06 |
| **NGC 1351** | 1428 ± 7 | 16.023859 | 1.30268E-06 |
| **NGC 1387** | 1208 ± 7 | 20.701286 | 1.20433E-06 |
| **NGC 1351A** | 1248 ± 8 | 22.672654 | 1.40781E-06 |
| **IC 1970** | 1167 ± 7 | 25.34644547 | 1.37617E-06 |
| **NGC 1406** | 974 ± 8 | 33.077567 | 1.25102E-06 |

**Table9 :** Perpendicular distances of the companion galaxies to the reference plane ($P_3$).

| Galaxy name | D⊥ (Mpc) | D⊥/L | Coplanaire à 10 % ? |
|---|---|---|---|
| **NGC 1399** | 0.000 | 0.00 % | yes |
| **ESO 548- G 063** | 0.6697 | 4.29 % | Yes |
| **NGC 1377** | 0.4462 | 2.86% | Yes |
| **NGC 1351** | 0.6735 | 4.31 % | yes |
| **NGC 1387** | 0.000 | 0.00 % | yes |
| **NGC 1351A** | 0.5765 | 3.69% | yes |
| **IC 1970** | 0.000 | 0.00 % | Yes |
| **NGC 1406** | 0.2564 | 1.64% | yes |

### 6.3.3. Velocity–Distance Relation and Orbital Interpretation

For the finalized NGC 1399 galaxy system, the velocity–distance relation was constructed (Figure 6) by plotting the CMB-frame velocities of the companion galaxies as a function of their distances ($D_{1-2}$) from the central elliptical galaxy. Within the proposed framework, these velocities are interpreted as proxies for the large-scale orbital motion of the companion galaxies around NGC 1399.

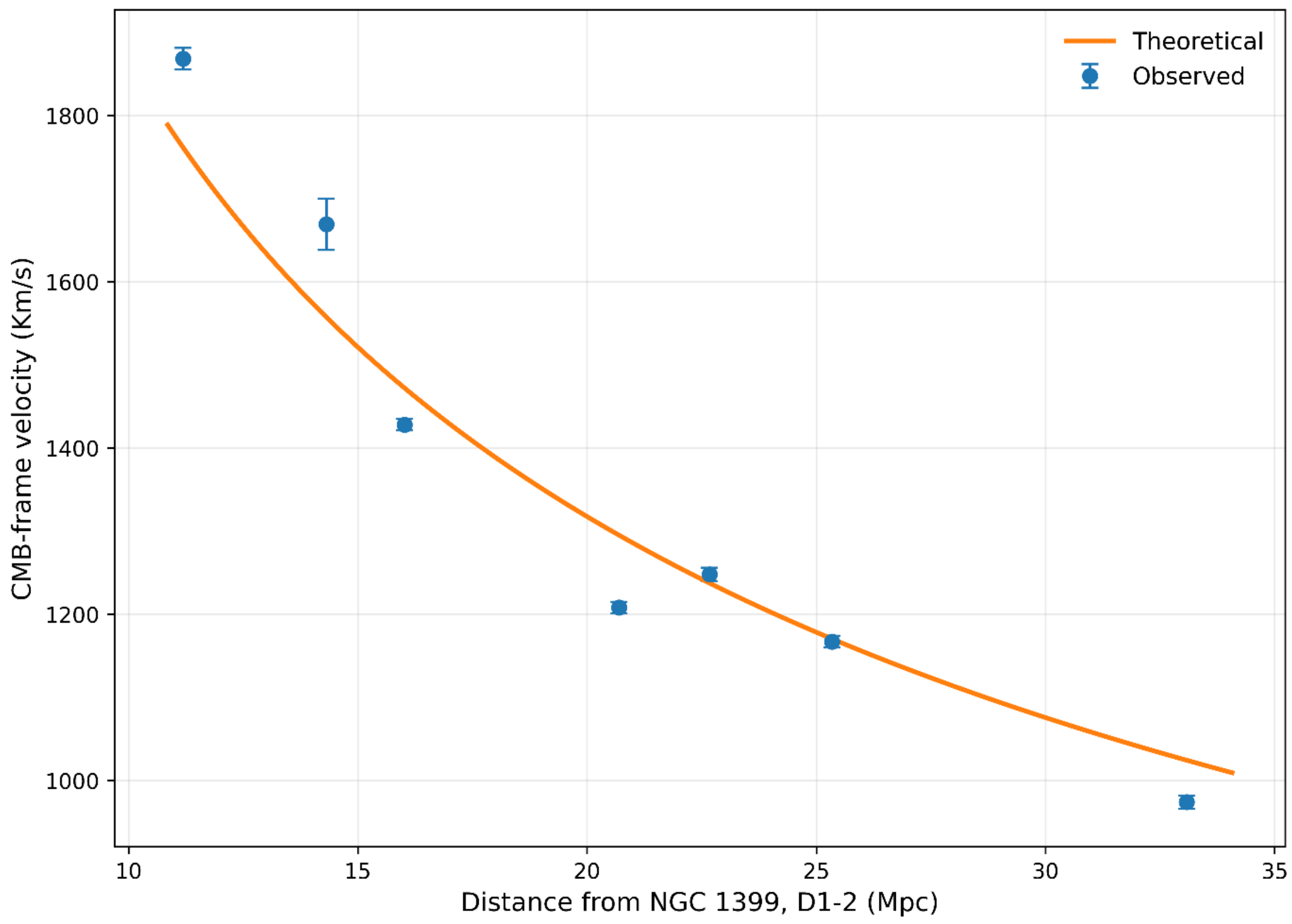

**Fig. 6.** Velocity–distance relation of the NGC 1399 system

#### 6.3.3.1. Interpretation of the velocity–distance relation of the NGC 1399 system

The observed velocities in figure6, exhibit a clear overall decrease with increasing distance, defining a coherent declining velocity–distance relation over the investigated radial range.

The theoretical profile, calculated using the photometric mass of NGC 1399, ($M_{phot}$ (NGC 1399) =7.74 x$10^{41}$ kg), and the mean effective gravitational parameter ($G_{eff}$ =1.38404 x$10^{-6}$ $m^3kg^{-1}s^{-2}$), shows the same overall Keplerian-like decline. Although some deviations are present for individual galaxies, the observed and theoretical profiles display a consistent global radial trend.

Together with the previously established coplanar configuration, this decreasing velocity–distance relation is consistent with the interpretation of NGC 1399 as the dominant central gravitational component of an organized orbital system. Given the limited sample size, the result should be regarded as evidence for consistency with the proposed orbital framework rather than as a definitive demonstration of an exact Keplerian relation.

### 6.4. Synthesis and Global Interpretation

The three galaxy systems investigated in this work exhibit a coherent combination of kinematic and geometric properties. In each system, the selected non-active companion galaxies show a declining velocity–distance relation that is qualitatively consistent with orbital motion in a gravitational field dominated by a central giant elliptical galaxy. The spatial analysis further indicates that the selected companions form approximately coplanar configurations according to the adopted ($D_{\perp}/L$) criterion.

A particularly notable result is the similarity of the mean effective gravitational parameters derived independently for the three systems. The mean values of $G_{eff}$ are (1.78935 x$10^{-6}$ , 1.4276 ×$10^{-6}$ and 1.38404×$10^{-6}$) $m^3$ $kg^{-1}$ $s^{-2}$ for M87, M60, and NGC 1399, respectively. Despite the fact that these systems are spatially distinct and are not members of the same galaxy cluster, the derived values remain within a relatively narrow range. This recurrence of a similar ($G_{eff}$ ) scale may indicate a common dynamical behavior among giant-elliptical-dominated systems within the framework proposed here. At this stage, however, this similarity should be regarded as an empirical result rather than as evidence for a universal modification of the gravitational constant.

Taken together, the recurrence of declining velocity–distance profiles, approximate coplanarity, and comparable ($G_{eff}$) values across three independent systems provides a consistent phenomenological basis for the proposed orbital interpretation. In this framework, giant elliptical galaxies are treated as dominant central gravitational components, while the surrounding non-active galaxies are interpreted as members of an organized orbital system. The results therefore suggest that the dynamics of these systems may be described, at least to first order, by a central-mass orbital framework based on classical gravitational dynamics.

An additional aspect further reinforces the conceptual analogy with the Solar System. Several of the dominant giant elliptical galaxies considered in this study, including M87 and NGC 1399, host active galactic nuclei that produce radio jets. Although these jets are not expected to influence the orbital dynamics of the surrounding galaxies, they provide an interesting phenomenological parallel with the Sun, whose magnetic activity is manifested through prominences and coronal mass ejections. This parallel is intended solely as an illustrative analogy and does not constitute a dynamical assumption of the proposed model.

In the following section, these findings are compared with the predictions of the ΛCDM and MOND paradigms, in order to assess the scope, strengths, and limitations of the proposed orbital framework within the broader context of galactic dynamics.

## 7. COMPARISON WITH ΛCDM AND MOND MODELS

To place the proposed orbital framework within the broader context of galactic dynamics, it is instructive to compare its underlying assumptions with those of the two principal paradigms currently used to interpret galaxy kinematics: the ΛCDM (Lambda Cold Dark Matter) cosmological model and the MOND (Modified Newtonian Dynamics) framework.

Within the ΛCDM paradigm, the observed kinematics of galaxies are interpreted as the consequence of gravitational potentials dominated by extended dark matter halos. In contrast, MOND explains many galactic rotation curves by modifying Newtonian dynamics below a characteristic acceleration scale rather than introducing non-baryonic dark matter.

The framework proposed in this work follows a fundamentally different approach. Rather than explicitly introducing a dark-matter component or modifying the laws of gravity, the present framework reinterprets the dynamical organization of a specific class of galaxy systems. In elliptical-dominated groups, the observed velocity–distance relations are interpreted as the natural consequence of orbital motion of companion galaxies around a dominant central giant elliptical galaxy.

Table 10 summarizes the fundamental assumptions, physical mechanisms, domains of applicability, strengths, and limitations of the three approaches. This comparison illustrates that the proposed framework does not introduce new particles or modify gravitational laws, but instead provides an alternative dynamical interpretation for a specific class of galaxy systems.

**Table 10.** Conceptual comparison between the ΛCDM model, MOND, and the proposed orbital framework.

| **Feature** | **ΛCDM** | **MOND** | **Proposed Orbital Framework** |
|---|---|---|---|
| **Primary hypothesis** | Galaxies are embedded in massive dark matter halos. | Newtonian dynamics is modified below a critical acceleration ($a_0 \approx 1.2\times10^{-10}$ ms$^{-2}$). | Giant elliptical galaxies act as dominant gravitational centers around which companion galaxies follow orbital trajectories. |
| **Origin of the observed kinematics** | Gravitational potential includes a dominant dark-matter component.. | Modified gravitational dynamics in the low-acceleration regime. | Classical Newtonian orbital motion within a centrally dominated gravitational potential. |
| **Physical mechanism** | Additional cold dark matter component contributes to the gravitational potential | Modification of the effective gravitational response relative to Newtonian dynamics | Reinterpretation of the spatial geometry and dynamical organization of selected galaxy systems. |
| **Kinematic interpretation** | Approximately flat galaxy rotation curves are explained through extended gravitational potentials including dark matter halos. | Approximately flat galaxy rotation curves are reproduced through modified dynamics at low accelerations. | Decreasing velocity–distance relations in selected giant-elliptical-dominated galaxy systems. |
| **Primary domain of application** | Galaxies, galaxy clusters, and cosmology. | Primarily individual galaxies; less successful for clusters and cosmology. | Giant-elliptical-dominated galaxy systems with approximately coplanar selected companions. |
| **Newtonian gravity** | Preserved, with an additional dark-matter component. | Modified below the critical acceleration ($a_0$). | Preserved; a system-specific effective dynamical parameter ($G_{eff}$) is introduced within the orbital formulation. |
| **Additional ingredients required** | Cold dark matter component. | Characteristic acceleration scale and interpolation prescription | No explicit dark-matter component or modification of Newtonian dynamics is introduced in the present formulation; the framework relies on the adopted orbital interpretation of the system geometry. |
| **Main strength** | Successfully accounts for large-scale structure formation and a broad range of cosmological observations. | Successfully reproduces many galactic rotation curves without invoking dark matter. | Provides a purely Newtonian interpretation of the observed kinematics in the selected galaxy groups. |
| **Main limitation** | The nature of dark matter remains unknown and a direct particle detection has not yet been achieved.. | Its application to galaxy clusters and cosmology remains challenging. | Its broader applicability requires testing with larger, independently selected samples and further numerical and observational investigation. |

Although ΛCDM remains the dominant cosmological framework because of its success in explaining large-scale structure formation and a wide range of cosmological observations, a direct detection of dark matter particles has not yet been achieved despite extensive experimental efforts. Likewise, MOND successfully reproduces many observed galactic rotation curves, while its application to galaxy clusters and several cosmological observations remains challenging. These considerations motivate the exploration of complementary dynamical interpretations for specific classes of galaxy systems.

The proposed orbital framework is not intended as a universal alternative to either ΛCDM or MOND. Rather, it provides a complementary interpretation for the giant-elliptical-dominated systems investigated here. In this framework, selected non-active companion galaxies are treated as members of an organized orbital configuration around a dominant central giant elliptical galaxy. Their decreasing velocity–distance relations are interpreted within classical Newtonian orbital dynamics, with a system-specific effective gravitational coupling ($G_{eff}$).

The principal distinction from the conventional frameworks summarized in Table 10 therefore concerns the assumed dynamical organization of the system: ΛCDM attributes the observed kinematics to gravitational potentials containing extended dark-matter components, MOND modifies the effective gravitational response in the low-acceleration regime, whereas the present framework emphasizes centrally dominated orbital motion of selected companions.

An additional result is the relatively narrow range of the mean ($G_{eff}$) values obtained for the three independent systems studied. Although these values are system-specific, their similarity suggests a recurring effective dynamical scale that warrants further investigation. This should be regarded as an empirical finding of the present sample rather than as evidence for a universal modification of Newtonian gravity. Broader applicability of the framework will require testing across a larger number of giant-elliptical-dominated systems.

## 8. LIMITATIONS OF THE PRESENT FRAMEWORK

The present results provide a consistent basis for the proposed orbital interpretation of giant-elliptical-dominated galaxy systems. As with any framework based on a specific sample and a set of simplifying assumptions, however, the scope of the present conclusions should be considered within the limits of the available data and methodology.

First, the present analysis focuses on three representative systems centered on M87, M60, and NGC 1399. Although these systems are independent and exhibit similar geometric and kinematic characteristics, analysis of a larger sample will be needed to determine how broadly the identified behavior extends to other giant-elliptical-dominated environments.

Second, the framework is deliberately focused on companion galaxies satisfying the selection criteria adopted in this study, including approximate spatial coplanarity with respect to the dominant central elliptical galaxy. This coplanarity criterion is an explicit structural assumption of the proposed framework and is motivated by the conceptual analogy with the approximately planar organization of planetary orbits in the Solar System. It is therefore used to identify galaxy systems in which an organized orbital configuration can be investigated. Accordingly, the framework does not seek to reproduce the full three-dimensional distribution of all galaxies in a group, but rather to characterize the large-scale orbital behavior of selected companions within the imposed structural configuration. Its applicability to systems with substantially different spatial or dynamical configurations remains to be investigated.

Third, the effective gravitational parameters ($G_{eff}$) derived here are model-dependent parameters defined within the proposed orbital framework. Their relatively similar values across the three independent systems constitute an empirical result of the present analysis, but should not be interpreted as a replacement for, or a violation of, the universality of Newton's gravitational constant. Further investigation with independent samples will be needed to determine whether the apparent recurrence of this effective dynamical scale persists more generally.

Finally, the present framework is formulated analytically and is based on a simplified central-mass orbital description. Dedicated numerical studies, including N-body simulations, would provide a useful means of examining the stability and long-term evolution of the proposed configurations and of assessing the effects of three-dimensional structure and non-ideal orbital motions. Improved distance determinations and additional kinematic constraints, including proper-motion measurements where feasible, would also help refine the orbital reconstruction and further test the broader applicability of the framework.

## 9. PERSPECTIVES AND FUTURE WORK

The orbital framework developed in this study provides several directions for further investigation. Extending the analysis to a larger selected sample of giant-elliptical-dominated systems will help determine whether the decreasing Keplerian-like velocity–distance relations identified here occur more broadly. Improved distance and velocity measurements, together with more complete three-dimensional mapping of the systems, will provide stronger constraints on the proposed orbital configurations and their spatial organization. Finally, dedicated N-body simulations could be used to examine the stability and long-term evolution of these configurations and to assess the possible roles of central mass and environment in shaping their dynamics.

## 10. CONCLUSION

In this study, we introduced an orbital framework for reinterpreting the large-scale kinematics of galaxy systems dominated by giant elliptical galaxies. In contrast to interpretations based primarily on rotationally supported configurations, the proposed approach retains classical Newtonian gravity and describes selected companion galaxies as gravitationally bound orbital bodies moving within the gravitational field of a dominant central giant elliptical galaxy.

Applied to three independent representative systems—M87, M60, and NGC 1399—the framework yields systematically decreasing velocity–distance relations that are qualitatively consistent with Keplerian-like orbital behavior. In systems characterized by a pronounced gravitational hierarchy and approximate coplanarity, the observed kinematics can be described within the proposed central-mass orbital framework without explicitly introducing an additional dark-matter component or a modification of Newtonian dynamics.

The analysis also yields system-specific effective gravitational parameters, ($G_{eff}$), whose values remain relatively coherent within each system. Importantly, despite the independence of the three systems, their mean ($G_{eff}$) values fall within a relatively narrow range, from approximately ($1,38404\times10^{-6}$ $m^3$ $kg^{-1}$ $s^{-2}$) to ($1,78935$ x$10^{-6}$ $m^3$ $kg^{-1}$ $s^{-2}$). This recurrence is noteworthy and may indicate a common effective dynamical scale among the giant-elliptical-dominated systems investigated here. However, ($G_{eff}$) should be regarded as a model-dependent effective parameter rather than as a replacement for the universal Newtonian gravitational constant. The origin of this relatively narrow range, and its possible dependence on the hierarchical and environmental properties of galaxy systems, warrants further investigation.

The proposed framework is not intended to replace established paradigms such as ΛCDM or MOND. Rather, it suggests that, in selected systems, some velocity–distance relations may be more naturally interpreted in terms of hierarchical orbital organization than collective rotational dynamics. By emphasizing the role of gravitational hierarchy, central mass dominance, and spatial organization, the present work provides a complementary perspective on the large-scale dynamics of giant-elliptical-dominated galaxy systems.

Future observational, statistical, and numerical studies involving larger and independently selected samples will be required to assess the robustness and broader applicability of this interpretation. In particular, improved distance measurements, additional kinematic constraints, and dedicated numerical modeling will help determine whether the patterns identified here recur systematically across similar environments.

## ACKNOWLEDGMENTS

The authors would like to express their sincere gratitude to the Algerian General Directorate of Scientific Research and Technological Development (DGRSDT) for its support. We also thank Professors José Gaite and Courtney Seligman, and Drs. Suvrat Rao, Mitchell F. Struble, Michal Michalowski, and Shreya Ojha, for their valuable discussions. We extend our sincere thanks to Dr. Marion Schmitz and the NED team for their valuable support. This work made use of the NASA/IPAC Extragalactic Database (NED), which is operated by the Jet Propulsion Laboratory, California Institute of Technology, under contract with the National Aeronautics and Space Administration.